\documentclass[pre,twocolumn,groupedaddress,showpacs,floatfix,superscriptaddress]{revtex4}
\usepackage{graphicx}
\usepackage{dcolumn}
\usepackage{amsmath}
\usepackage{bm}
\begin{document}

\title{Interaction of colloids with a nematic-isotropic interface}

\author{D.~Andrienko}

\affiliation{Max Planck Institute for Polymer Research, Ackermannweg
  10, 55128 Mainz, Germany}

\author{M.~Tasinkevych} 

\affiliation{Departamento de F\'\i sica da Faculdade de Ci\^encias and
  Centro de F\'\i sica Te\'orica e Computacional, Universidade de
  Lisboa, Avenida Professor Gama Pinto 2, P-1649-003 Lisboa Codex,
  Portugal}

\author{P.~Patr{\'\i}cio}

\affiliation{Departamento de F\'\i sica da Faculdade de Ci\^encias and
  Centro de F\'\i sica Te\'orica e Computacional, Universidade de
  Lisboa, Avenida Professor Gama Pinto 2, P-1649-003 Lisboa Codex,
  Portugal} 

\affiliation{Instituto Superior de Engenharia de Lisboa, Rua
  Conselheiro Em\'\i dio Navarro 1, P-1949-014 Lisboa, Portugal}

\author{M.~M.~Telo da Gama}

\affiliation{Departamento de F\'\i sica da Faculdade de Ci\^encias and
  Centro de F\'\i sica Te\'orica e Computacional, Universidade de
  Lisboa, Avenida Professor Gama Pinto 2, P-1649-003 Lisboa Codex,
  Portugal}

\date{\today}
\begin{abstract} 
  The Landau-de Gennes free energy is used to calculate the
  interaction between long cylindrical colloids and the
  nematic-isotropic (NI) interface. This interaction has two
  contributions: one is specific of liquid crystals and results from
  the deformation of the director field close to the particles or to
  the interface, while the other is generic and results from wetting
  and surface tension effects.
  
  Deep in the nematic phase the director field of long cylindrical
  colloids, with strong homeotropic anchoring, exhibits two
  half-integer defect lines. As the colloid moves towards the
  interface, the director configuration changes through a series of
  discontinuous transitions, where one or two of the defects are
  annihilated. In addition, the NI interface bends towards the colloid in
  order to minimize the elastic free energy in the nematic.  In the
  isotropic phase, the colloid is surrounded by a thin nematic layer that
  reduces the surface free energy under favorable wetting conditions.
  
  The interaction has a well-defined minimum near the interface. In
  this region the director and interfacial structures are complex and
  cannot be described analytically. Using the numerical results for
  the Landau-de Gennes free energy in the harmonic region, we obtained
  simple scaling laws for the (linear) force on the colloid.
\end{abstract}
\maketitle

\section{Introduction}

Colloidal dispersions are suspensions of solid or liquid particles in
a host fluid. The size of the particles can vary from a
few $\rm nm$ up to several $\mu \rm m$. Colloidal dispersions are
often found in long-lived metastable states, providing the basis for a
range of industrial applications, e.~g. paints, drugs, foods,
coatings, etc~\cite{russel.wb:1989.a}.

Colloidal dispersions in nematic liquid crystals form a special class
of colloids. The difference from ordinary colloids arises
from the long-range orientational order of the liquid crystal
molecules, described by the so-called nematic {\em director}.
Topological defects of the director field~\cite{lubensky.tc:1998.a},
additional long-range forces between colloidal
particles~\cite{ramaswamy.s:1996.a,poulin.p:1997.a,lubensky.tc:1998.a},
and, as a result, supermolecular
structures~\cite{lev.bi:1999.a,loudet.jc:2001.a} are phenomena
specific of colloidal nematics.

Colloidal nematics are usually prepared in the isotropic phase. To
prevent flocculation due to attractive van~der~Waals forces, colloidal
particles are treated to induce electrostatic or steric repulsive
interactions. After cooling to a temperature below the NI transition,
the colloids often segregate forming non-uniform clusters. In
general, topological defects stabilize the
dispersion~\cite{poulin.p:1997.a} although, in some cases, they may
help flocculation by producing additional attractive
interactions~\cite{tasinkevych.m:2002.a}.

The spatial distribution of colloidal particles is very sensitive to
the cooling rate and cooling
conditions~\cite{anderson.vj:2001.a,anderson.vj:2001.b}. Recent
results~\cite{west.jl:2002.a} revealed that the drag on colloids by a
NI interface plays an important role in the final spatial pattern. In
the experiment, large nematic and isotropic domains were separated by
moving NI interfaces. Depending on the cooling rate (velocity of the
interface) different structures were observed: a) cellular structures,
also reported
earlier~\cite{anderson.vj:2001.a,anderson.vj:2001.b,meeker.sp:2000.a},
with particle-free nematic domains separated by particle-rich regions;
b) striped structures, where the particle-rich regions are arranged in
a set of stripes; c) root-like structures. A major conclusion of the
study was that the motion of the particles may be controlled by their
interaction with the NI interface.

Here we reexamine this conclusion by analyzing in detail the
interaction between colloidal particles and a NI interface.  As a
first step we investigate the (static) Landau-de Gennes order
parameter distributions around colloidal particles in the vicinity of
NI interfaces. By calculating the corresponding free energies we
obtain explicit results for the forces.

For simplicity, we consider the interaction of a long cylindrical
colloid with a planar NI interface. We assume strong
homeotropic anchoring of the director at the particle's surface
everywhere and wetting boundary conditions (see
section~\ref{sec:theory_free_en}).  In the nematic phase, these
anchoring conditions give rise to a pair of $1/2$ strength defect
lines, that merge with the isotropic phase as the particle approaches
the NI interface, distorting the planar interfacial region on a scale
of the order of the colloidal radius.  As a result, the interface is
strongly bent around the colloidal particle on both sides of the
interface. The effective interaction is shown to be rather complex and
to exhibit a well defined minimum close to the interface. Near the
minimum, a scaling analysis of the numerical results for the free
energy yields the scaling form of the force on the particle.

The paper is organized as follows. The geometry, the model, and the
minimization technique are described in Sec.~\ref{sec:theory}. The
results for the director configurations and free energy are presented
in Sec.~\ref{sec:results}. A discussion of the major findings,
together with concluding remarks, appear in Sec.~\ref{sec:discussion}.

\section{Landau-de Gennes theory}
\label{sec:theory}
\subsection{Geometry}
\begin{figure}
\begin{center}
\includegraphics[width=6cm, angle=0]{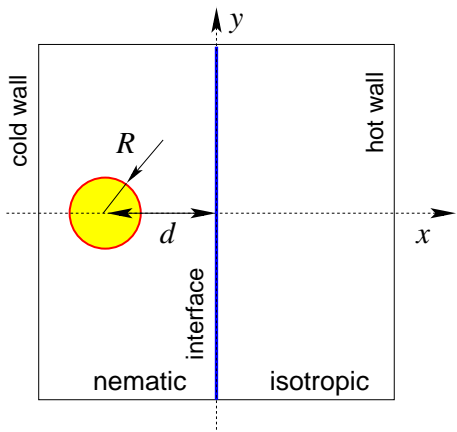}
\end{center}
\caption[Studied geometry]{
\label{fig:1} % fig1.eps
$xy$ cross-section, of size $L\times L$, of the system studied.}
\end{figure}

In this study we consider a geometry similar to that used in the
experiments~\cite{west.jl:2002.a}.  Briefly, one end of the sample is
placed in a hot and the other in a cold oven. The temperatures are
chosen such that the NI interface is in the gap between the ovens and
a constant temperature gradient is maintained throughout the sample.

A geometry mimicking the experimental setup is shown in
Fig.~\ref{fig:1}. A uniform temperature gradient is imposed along the
$x$ axis with the `hot' wall at $x = L/2$ and the `cold' wall at $x =
-L/2$, where $L$ is the size of the system. The director at the `cold'
wall is fixed either parallel or perpendicular to the wall. The order
parameter at the `hot' wall is set to zero. Note, that the position of
the interface, at $x=0$, is pinned by the temperature gradient.

A colloidal particle, which we take to be a long cylinder of radius
$R$, with the symmetry axis parallel to the $z$ axis, is immersed in
the nematic. Periodic boundary conditions are used in the $y$
direction.

\subsection{Tensor order parameter}
We use the Landau-de Gennes tensor order-parameter $Q_{\alpha \beta}$
formalism since it is free from {\em divergent} terms
due to the defect cores and it takes into account the {\em biaxiality}
that may occur in non-uniform
nematics~\cite{schopohl.n:1987.a,schopohl.n:1988.a,popanita.v:1997.a}.

Owing to the traceless, symmetric character of the tensor order
parameter it can be represented as~\cite{sen.ak:1987.a}
\begin{equation}
  Q_{\alpha \beta}({\bm r}) = 
   \frac{1}{2}Q(3n_{\alpha}n_{\beta} - \delta_{\alpha \beta}) 
  +\frac{1}{2} B( l_{\alpha}l_{\beta} - m_{\alpha}m_{\beta}),
\end{equation}
where the direction of maximal orientational order is given by the
director $\bm n$, $Q$ is the scalar order parameter, and the unit
vectors $\bm l, \bm n, \bm m$ form a local orthonormal triad.

Spatial non-uniformities that do not coincide with ${\bm n}$ break the
cylindrical symmetry of the average angular environment, and thus one
must allow for biaxiality, i.e. $B \ne 0$.

The translational symmetry of the interface along the $z$ axis and the
assumed homeotropic boundary conditions on the colloidal surface,
imply that the director is confined to the $xy$ plane. Then the vector
$\bm m$ may be chosen along the $z$ axis and the vector $\bm l$ is in
the $xy$ plane
\begin{eqnarray}
\bm n &=& (\cos \theta, \sin \theta, 0), \\
\nonumber
\bm l &=& (\sin \theta, -\cos \theta, 0), \\
\nonumber
\bm m &=& (0, 0, 1).
\end{eqnarray}

In this case, the tensor order parameter has three independent
components only
\begin{equation}
\displaystyle{
{\bm Q} = \left(
\begin{array}{ccc}
Q_{11} & Q_{12} & 0 \\
Q_{12} & Q_{22} & 0 \\
0      & 0      & -Q_{11} - Q_{22}          
\end{array}
\right)}
\label{eq:Q}
\end{equation}
In the following we will use this representation of the tensor order
parameter.

\subsection{Free energy}
\label{sec:theory_free_en}
The system is described by the Landau-de Gennes free
energy~\cite{degennes.pg:1995.a}
\begin{equation}
{\cal F}\{\bm Q\} = 
\int{(f_b + f_e)dV} + \int{f_s dS}
\label{eq:free_en}
\end{equation}
where $f_b$ is the bulk free energy density, $f_e$ is the elastic free
energy density and $f_s$ is the surface free energy. Within a
mesoscopic approach the minimum of the Landau-de Gennes functional
$\cal F$ gives the equilibrium value of the tensor order parameter.

Symmetry arguments yield for the local bulk free energy
density~\cite{stephen.mj:1974.a,degennes.pg:1995.a}
\begin{equation}
f_b = a {\rm Tr} {\bm Q^2} -
      b {\rm Tr} {\bm Q^3} +
      c \left[{\rm Tr} {\bm Q^2}\right]^2,
\label{eq:f_b}
\end{equation}
where $a$ is assumed to depend linearly on the temperature, while the
positive constants $b,c$ are taken temperature independent.

It is convenient to scale out the variables by defining
\begin{eqnarray}
\tilde{Q}_{ij} = 6c/b Q_{ij}, \\ \nonumber
\tilde{f}_b = 24^2c^3/b^4 f_b.
\end{eqnarray}
It will be understood that such scaling has been carried out, and we
shall omit the overbars in the text below.

We also introduce a dimensionless temperature $\tau$ by defining
\begin{equation}
a = \tau b^2/24c.
\end{equation}

For a {\em uniform uniaxial} nematic ($Q_{11} = Q$, $Q_{22} = Q_{33} =
-1/2Q$) the free energy (\ref{eq:f_b}) takes the form
\begin{equation}
f_{b} = \tau Q^2 - 2Q^3 + Q^4.
\end{equation}
The nematic state is stable when $\tau < 1$ with a degree of orientational 
order given by
\begin{equation}
Q_b = \frac{3}{4}\left(1+\sqrt{1-\frac{8}{9}\tau} \right).
\end{equation} 
We modeled the temperature gradient in the $x$ direction by assuming
that $\tau$ depends on the $x$ coordinate
\begin{equation}
\tau = \tau_c\left(1 + \alpha \frac{2x}{L}\right),
\label{eq:grad}
\end{equation}
where $\tau_c = 1$ is the NI transition temperature.
Equation~(\ref{eq:grad}) implies that the NI transition occurs at $x =
0$, {\sl i.e.}, the interface is in the middle of the cell and is
parallel to the $y$ axis.

The elastic free energy density can be written
as~\cite{stephen.mj:1974.a}
\begin{equation}
f_e = \frac{1}{2} L_1 
\frac{\partial Q_{ij}}{\partial x_k} \frac{\partial Q_{ij}}{\partial x_k} +
\frac{1}{2} L_2
 \frac{\partial Q_{ij}}{\partial x_j} \frac{\partial Q_{ik}}{\partial x_k},
\end{equation}
where the constants $L_1$ and $L_2$ are related to Frank-Oseen elastic
constants by $K_{11} = K_{33} = 9Q_b^2(L_1 + L_2/2)/2$ and $K_{22} =
9Q_b^2L_1/2$ and $Q_b$ is the bulk nematic order parameter. The
sign of $L_2$ defines the preferred orientation of the director at the
NI interface. $L_2>0$ ($L_2<0$) favors planar (perpendicular)
anchoring~\cite{degennes.pg:1971.a}.
 
We assumed strong homeotropic anchoring of the director at the
colloidal surface. This is valid if the anchoring parameter $WR/K >>
1$, where $W$ is the anchoring energy of the surface, and holds for
large colloidal particles and/or anchoring strengths. We also assumed
that the nematic at the particle surface is uniaxial with a scalar
order parameter $Q_{\rm s} =1$, independent of $x$. Under these
conditions, a planar interface is wetted by the nematic phase as the
temperature is lowered through the NI transition.

\subsection{Minimization procedure}
The equilibrium distribution of the tensor order parameter $Q_{ij}$ is
obtained by minimizing the free energy functional~(\ref{eq:free_en})
numerically using finite elements with adaptive meshes. During the
minimization the square integration region $L \times L$ was
triangulated using the $\rm BL2D$ subroutine~\cite{george.pl:1998.a}.
The functions $Q_{ij}$ are set at all vertices of the mesh and are
linearly interpolated within each triangle. The free energy is then
minimized using the conjugate gradients method~\cite{press.wh:1992.a}
under the constraints imposed by the boundary conditions.

A new adapted mesh is generated iteratively from the previous
minimization. The new local triangle sizes are calculated from the
variations of the free energy, in order to guarantee a constant
numerical weight for each minimization variable. The final meshes,
with a minimal length of $\sim 10^{-2}$, had $\sim 10^5$
minimization variables. Lengths are in units of the diameter of the
colloidal particle, $2R = 1$.

We used $\xi^2=L_1 8c/b^2= 0.001$ that corresponds to a nematic
correlation length $\xi\approx 0.03$, which is larger
than the smallest mesh length. The system size was either $L=12$ or
$L=20$ and the temperature gradient was set at $\alpha = 0.025$.

In order to obtain both stable and metastable configurations, we used
different types of initial conditions including flat and curved NI
interfaces.

\section{Results}
\label{sec:results}
\subsection{Free interface}
First, we analyze the tensor order parameter of a free NI interface,
without colloidal particles and temperature gradients, for a system
with $L_2/L_1 = 2$. For this ratio of the elastic constants the
interface favors planar anchoring of the director (molecules parallel
to the interface)~\cite{degennes.pg:1971.a}.  The order parameter is
fixed at the cell boundaries: $Q(x=-L/2)=1$, $Q(x=L/2)=0$ and the
director is aligned along the $y$ axis, parallel to the interface.

\begin{figure}
\begin{center}
\includegraphics[width=8cm]{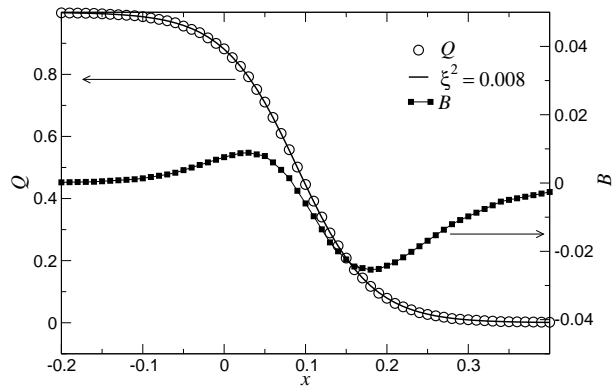}
\end{center}
\caption{
  Order parameter $Q$ and biaxiality $B$ of a free NI interface.
  Circles: order parameter, results of the numerical minimization;
  solid line: fit to Eq.~(\ref{eq:ord_interface}). Squares:
  biaxiality, numerical results. Note the different scales and offset
  for the order parameter and biaxiality profiles.  }
\label{fig:interface}
\end{figure}

The order parameter and biaxiality profiles are shown in
Fig.~\ref{fig:interface}. As expected, the order parameter changes
from the bulk value of the nematic phase, $Q_b = 1$, to the bulk value
of the isotropic phase, $Q_{\rm iso} = 0$. The width of the interface
is of the order of the nematic correlation length. The nematic and
isotropic phases are uniaxial, with zero biaxiality. The interfacial
region, however, is slightly biaxial.

Note, that the analytic solution of the corresponding variational
problem does not exist, even in the one-dimensional case, when the
tensor order parameter depends only on the $x$ coordinate. However,
one may use the de~Gennes Ansatz for the order parameter profile,
which neglects biaxiality~\cite{degennes.pg:1971.a},
\begin{equation}
Q = \frac{1}{2}\left( 1-\tanh\frac{x}{\zeta} \right),
\label{eq:ord_interface}
\end{equation}
where $\zeta$ is the nematic correlation length when the director is
parallel to the interface
\begin{equation}
\zeta^2 = \xi^2 \left( 6 + \frac{L_2}{L_1}\right).
\label{eq:ord_interface_theory}
\end{equation}
A fit of the results of the numerical minimization to
Eq.~(\ref{eq:ord_interface}) yields $\zeta^2 \approx 0.00807$ while
Eq.~(\ref{eq:ord_interface_theory}) gives $\zeta^2 = 0.008$, {\sl
  i.e.}, the agreement is very good.  The small difference is
attributed to the biaxiality of the
interface~\cite{popanita.v:1997.a}, not taken into account by the
Ansatz (\ref{eq:ord_interface}). Note that we decreased the minimal
mesh length down to $10^{-3}$ in order to increase the accuracy of
the results.

We also found that the surface tension is a function of the polar
angle of the director with the interface normal.  As a result, the
interface has an easy axis with corresponding anchoring energy. We
found planar interfacial anchoring (easy axis parallel to the
interface) for $L_2 > 0$. $L_2 < 0$ results in homeotropic interfacial
anchoring (easy axis normal to the interface) in agreement with
previous results~\cite{degennes.pg:1971.a,popanita.v:1997.a}.

We studied the interaction of a colloid with the NI interface under
various anchoring conditions of the director at the interface: i)
director parallel to the interface, $L_2 > 0$; ii) perpendicular to
the interface, $L_2 < 0$; iii) director tilted with respect to the
interface, due to a mismatch between interfacial anchoring and the
alignment at the cell boundaries.

\subsection{Planar interfacial anchoring}
\label{sec:results_planar}
The anisotropy of the elastic constants is $L_2/L_1 = 2$, favoring
director alignment parallel to the NI interface. The director at the
`cold' wall is fixed parallel the $y$ axis, which is also parallel to
the NI interface.

\begin{figure}
\begin{center}
\includegraphics[width=8cm]{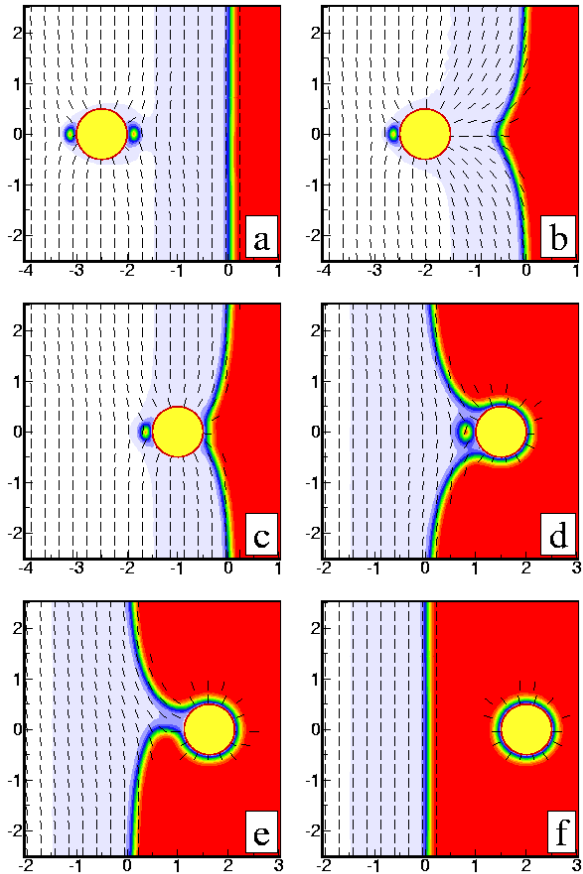}
\end{center}
\caption{
  Order parameter and director maps for colloids at a distance, $d$,
  from the NI interface: (a) $d = -2.5$; (b) $d=-2$; (c) $d=-1$; (d)
  $d=1.5$; (e) $d=1.625$; (f) $d=2$. Red (dark) corresponds to the isotropic
  phase with $Q=0$, and white to the nematic phase with $Q \approx 1$.
  System size $L = 12$, anisotropy of the elastic constants
  $L_2/L_1=2$.}
\label{fig:order_par}
\end{figure}

Typical order parameter and director maps are shown in
Fig.~\ref{fig:order_par}. The strong anchoring of the director at the
colloidal surface yields two half-integer defect lines in the nematic,
far from the interface, to ensure that the topological charge of the
system is zero (see Fig.~\ref{fig:order_par}a). The director
distortion vanishes very rapidly in the nematic phase, and the cores
of the defects extend over a few nematic correlation lengths, in line
with previous
studies~\cite{andrienko.d:2002.b,andrienko.d:2003.a,tasinkevych.m:2002.a}.
At these distances there is no interaction between the colloid and the
interface: the director at the interface is uniform, the interface is
flat and it is pinned at $x=0$, where the nematic and isotropic phases
coexist ($\tau = 1$).

\begin{figure}
\begin{center}
\includegraphics[width=8cm]{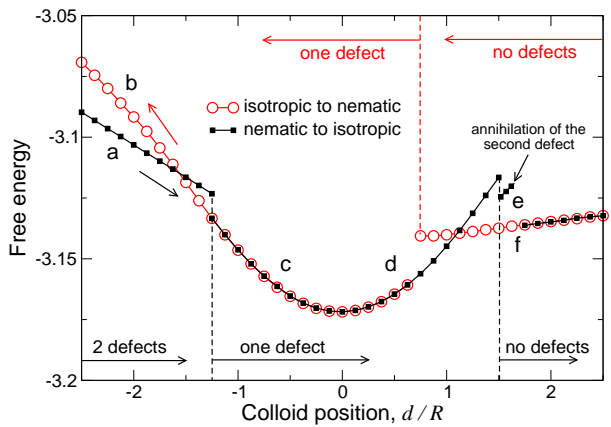}
\end{center}
\caption{
  Free energy $\cal F$ as a function of the distance of a colloidal
  particle from the NI interface. Both metastable and stable solutions
  are shown. Circles: particle moving from the isotropic to the
  nematic phase; squares: particle moving from the nematic to the
  isotropic phase.  System size $L = 12$, anisotropy of the elastic
  constants $L_2/L_1=2$. Letters correspond to the director
  configurations shown in Fig.~\ref{fig:order_par}.}
\label{fig:fen_par}
\end{figure}

On reducing the colloidal distance from the interface, the defect
closest to the interface merges (discontinuously) with the isotropic
phase (Fig.~\ref{fig:order_par}b,c). The interface bulges towards the
colloid to accommodate the isotropic phase, where the defect core
disappeared.  Note that the anchoring at the NI interface is planar
and follows the interfacial curvature, except in the region where the
defect used to be. Here the director tilts and the interfacial
anchoring becomes homeotropic, see Fig.~\ref{fig:order_par}b.

As the colloid moves further into the isotropic phase, it is wrapped
by the NI interface that forms a nematic `cavity' around the colloid
(Fig.~\ref{fig:order_par}d). The second defect is still present on the
nematic side.  At a certain point, this configuration becomes
metastable, and eventually annihilation of the second defect occurs.
This is accompanied by a symmetry breaking transition: the cavity is
no longer symmetric under $y\to-y$ reflexion
(Fig.~\ref{fig:order_par}e illustrates one of the two possible
configurations).  These configurations with broken symmetry are always
metastable.

Finally, deep in the isotropic phase, the colloid is wrapped by a thin
layer of nematic phase due to the (wetting) boundary conditions at the
colloidal surface ($Q_s=1$), Fig.~\ref{fig:order_par}f.

To investigate the nature of the structural transitions between
different director field configurations, we plot the free energy as a
function of the particle position in Fig.~\ref{fig:fen_par}.

Deep in the isotropic phase, the configuration of
Fig.~\ref{fig:order_par}f is the only stable one. There is a small
force acting on the particle in the direction opposite to the
temperature gradient. This force is due to an increase in the free
energy of the nematic layer, as the particle moves into a region at
higher temperature.

Moving from the isotropic into the nematic phase, this configuration
becomes first metastable, and then unstable. At a certain point, the
configuration with a nematic `cavity' in the isotropic phase,
Fig.~\ref{fig:order_par}d, becomes stable.

Finally in the nematic phase, the configuration with one defect
becomes metastable, and then unstable. The second defect appears, as
shown in Fig.~\ref{fig:order_par}a. At this point the interaction of
the particle with interface almost vanishes. The small negative slope
of the free energy, giving rise to a force along the temperature
gradient, is due to the temperature dependence of the elastic energy,
that increases as the particle moves towards the cold wall.

The jumps in the free energy reveal discontinuous (or first-order)
instabilities between the physical director configurations. The
equilibrium position, {\sl i.e.}, the position where the force on the
colloid vanishes, is near the interface, at $x = 0$.

\subsection{Homeotropic interfacial anchoring}

Liquid crystals with negative elastic constant anisotropy, $K_{22} >
K_{11},K_{33}$, are modeled by $L_2/L_1 < 0$
~\cite{degennes.pg:1995.a}.  In these systems the NI interface favors
homeotropic anchoring, {\sl i.e.}, alignment of the molecules
perpendicular to the interface~\cite{degennes.pg:1971.a}.

To study this case, we fixed the director at the cold wall along the
$x$ axis and used for the elastic constants $L_2/L_1
= -0.5$. For this set of parameters the director is parallel
to the $x$ axis, or perpendicular to the interface.

\begin{figure}
\begin{center}
\includegraphics[width=8cm]{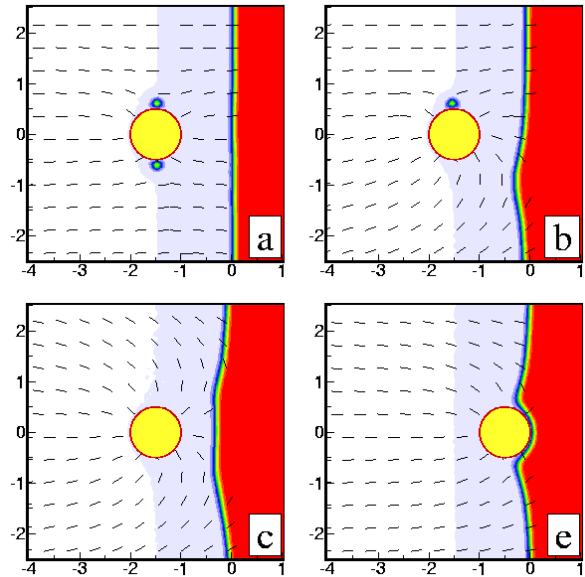}
\end{center}
\caption{
  Order parameter and director maps for colloids at a distance, $d$,
  from the NI interface: (a,b,c) $d = -1.5$; (e) $d=-0.5$. System size
  $L = 20$, anisotropy of the elastic constants $L_2/L_1=-0.5$. }
\label{fig:order_neg}
\end{figure}

\begin{figure}
\begin{center}
\includegraphics[width=8cm]{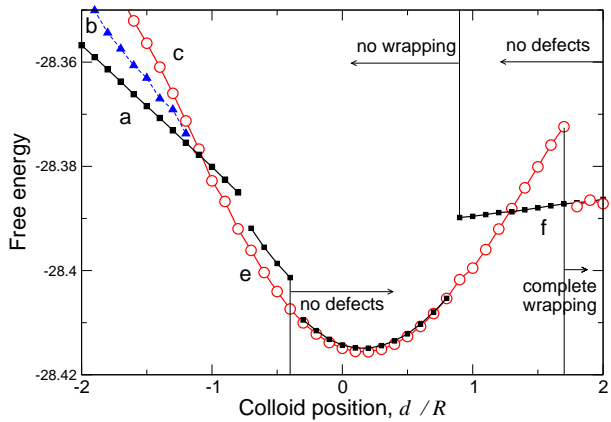}
\end{center}
\caption[Free energy]{
  Free energy $\cal F$ as a function of the distance of a colloidal
  particle from the NI interface.  Squares: initial condition with a
  `flat' interface; circles: initial condition with a `curved'
  interface; triangles: `curved' interface with the initial director
  tilted at $\pi/4$ (metastable configuration).  System size $L = 20$,
  anisotropy of the elastic constants $L_2/L_1=-0.5$.  Letters
  correspond to the director configurations shown in
  Fig.~\ref{fig:order_neg}.}
\label{fig:fen_neg}
\end{figure}

The order parameter and director maps are shown in
Fig.~\ref{fig:order_neg}. Since the director is perpendicular to the
interface, the defect lines are in the plane parallel to the
interface, Fig.~\ref{fig:order_neg}a. In this geometry both defects
merge with the isotropic phase very quickly, when the particle is
still on the nematic side of the interface, Fig.~\ref{fig:order_neg}e.
Subsequently, the particle is surrounded by a thin layer of nematic
phase until the fully wrapped state in the isotropic phase obtains,
Fig.~\ref{fig:order_par}f.

Deep in the nematic phase, we have also found a configuration with
only one defect, Fig.~\ref{fig:order_neg}b. However, analysis of the
free energy, Fig.~\ref{fig:fen_neg}, shows that this configuration is
always metastable.

\subsection{Director tilted at the NI interface}

Another situation, often found in experiments, corresponds to tilted
NI interfaces. However, the (simple) Landau-de Gennes theory fails to
describe this interfacial anchoring and higher order gradients or more
sophisticated density-functional theories are required in order to
describe tilt at NI interfaces~\cite{delrio.em:1995.a}. In finite
cells, on the other hand, tilted interfaces may result from
mismatching boundary conditions at the cell boundaries and the NI
interface proper. Indeed, in a planar cell, one may have the director
aligned parallel to the bounding plates, say, along the $x$ axis and a
NI interface in the $zy$ plane providing planar anchoring along the
$z$ axis. This is the so-called hybrid, or
$\pi$-cell~\cite{bos.pj:1984.a,sparavigna.a:1994.a} where the director
bends in the bulk to match the boundary conditions; as a result there
is a bulk splay-bend deformation even in the absence of colloidal
particles.

For this geometry we used $L_2/L_1 = 2$ and fixed the director at the
`cold' wall parallel to the $x$ axis or perpendicular to the NI
interface. The system size is $L=20$.

\begin{figure}
\begin{center}
\includegraphics[width=8cm]{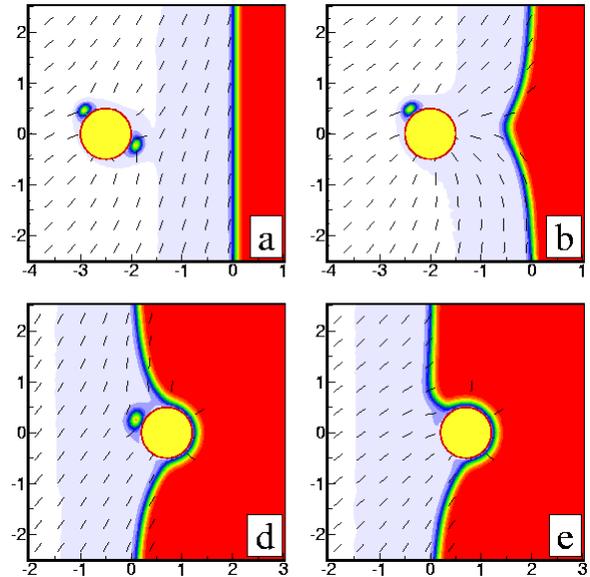}
\end{center}
\caption{
  Order parameter and director maps for colloids at a distance, $d$,
  from the NI interface: (a) $d = -2.5$; (b) $d=-2$; (d,e). Red (dark)
  corresponds to the isotropic phase with $Q=0$ and white to the
  nematic phase with $Q \approx 1$. System size $L = 20$, anisotropy
  of the elastic constants $L_2/L_1=2$. }
\label{fig:order_perp}
\end{figure}

\begin{figure}
\begin{center}
\includegraphics[width=8cm]{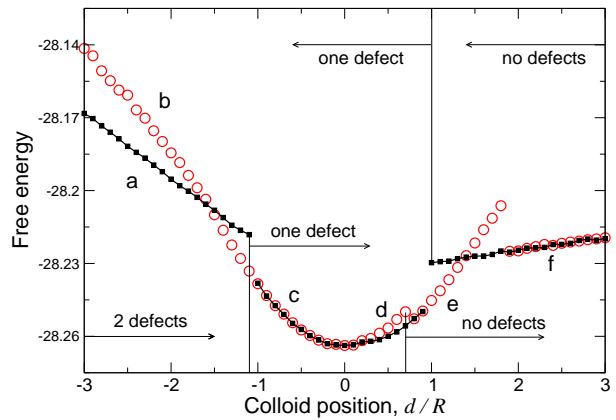}
\end{center}
\caption{
  Free energy $\cal F$ as a function of the distance of a colloidal
  particle from the NI interface. Squares: initial condition with a
  `flat' interface; circles: initial condition with a `curved'
  interface. System size $L = 20$, anisotropy of the elastic constant
  $L_2/L_1=2$. The director is tilted at the NI interface. Letters
  correspond to the director configurations shown in
  Fig.~\ref{fig:order_perp}. }
\label{fig:fen_perp}
\end{figure}

The order parameter and director maps, shown in
Fig.~\ref{fig:order_perp}, are qualitatively similar to those
described in section~\ref{sec:results_planar}, where the director is
parallel to the cold wall and the interface.  The tilt of the defects
is due to the tilted nematic director, Fig.~\ref{fig:order_perp}a.
Otherwise, the same configurations are present: with two defects, deep
in the nematic phase (Fig.~\ref{fig:order_perp}a); with one defect,
close to the interface (Fig.~\ref{fig:order_perp}b,d); without
defects, also close to the interface (Fig.~\ref{fig:order_perp}e);
without defects, deep in the isotropic phase (similar to the
configuration shown in Fig.~\ref{fig:order_par}f).

However, a careful analysis of the free energy, which is shown in
Fig.~\ref{fig:fen_perp}, reveals that the structural transitions are
different.  The configuration with one defect
(Fig.~\ref{fig:order_perp}d) becomes metastable in the isotropic side
of the interface; the defect-free configuration
(Fig.~\ref{fig:order_perp}e), that was metastable previously, is now
stable. A physical explanation is simple: because of the tilt, the
remaining defect is closer to the interface and it is more easily
annihilated.  The bending of the interface is also weaker.

\section{Discussion and conclusions}
\label{sec:discussion}
We end with a brief discussion of the implications of our work on the
drag of colloidal particles by the NI interface.

The free energy, Figs.~\ref{fig:fen_par},~\ref{fig:fen_neg},
and~\ref{fig:fen_perp} reveals that close to the minimum $\cal F$ is a
quadratic function of the position of the colloid and thus,
in this region, the force is proportional to the separation,
$d$.

\begin{figure}
\begin{center}
\includegraphics[width=8cm]{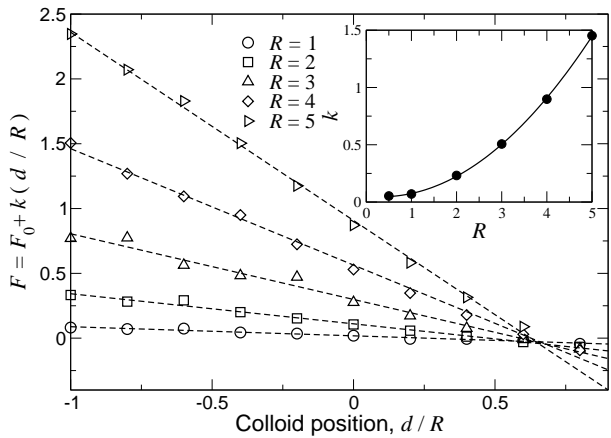}
\end{center}
\caption{
  Force on the particle as a function of the particle position. Inset:
  strength of the force as a function of the colloidal size and fit to
  $k \propto R^2$.}
\label{fig:force}
\end{figure}

We have calculated the force $F = -\partial {\cal F} / \partial d$, as
a function of $d$, for several colloidal sizes, in the geometry with
planar interfacial anchoring, as in the
section~\ref{sec:results_planar}.  The resulting curves, together with
the fit to the following dependence
\begin{equation}
F = F_0 + k\frac{d}{R}
\end{equation}
are shown in Fig.~\ref{fig:force}. The inset of Fig.~\ref{fig:force}
shows the dependence of the force strength $k$ on the particle size.
We find that, to a good approximation, $k \propto R^2$.

It is interesting to compare this result to the scaling arguments used
in Ref.~\cite{west.jl:2002.a}. If we neglect phenomena related to the
creation and annihilation of defects, and treat the interface as 
sharp, the only parameters in the problem are the
radius and length of the colloid, $R$ and $L$ respectively,
the elastic constant $K$, and the surface tension $\sigma$.  The
appropriate combinations of these parameters with the dimensions of
force are $KL/R$ and $\sigma LR$, where we took into account that for
a particle of length $L >> R$ the force is proportional to $L$.
Therefore, the elastic and the surface tension contributions to the
total force can be written as $F_{\rm e} = KL/R f_{\rm e}(d/R)$,
$F_{\rm s} = L \sigma f_{\rm s}(d/R)$, where $f_{\rm e}$ and $f_{\rm
  s}$ are dimensionless functions of the penetration depth, $d/R$.  It
is clear that neither of these forces gives the observed $\propto R^2$
dependence on the particle size: the elastic force is inversely
proportional to $R$, while the contribution due to the surface tension
does not depend on the particle size at all. One reason for this
discrepancy (apart from the presence of defects) is the complex
structure of the interface, that bends forming a cavity to wrap the
colloid.

Another interesting conclusion is the asymmetry of the free energy
profile as a function of the colloid position: when the particle moves
from the nematic to the isotropic phase, the metastable phase with a
nematic `cavity' (Fig.~\ref{fig:order_par}d) becomes unstable much
faster than the corresponding metastable state with one or no defects
(Figs.~\ref{fig:order_par}b), when the particle moves from the
isotropic to the nematic phase.  Therefore, the interface drags the
particle more efficiently when the colloid moves from the nematic to
the isotropic side, owing to the presence of a higher energy
barrier. This effect has been observed
experimentally~\cite{west.jl:2002.a}.

We also note that wrapping a colloidal particle by the NI interface
resembles the process of wrapping a colloidal particle by a
membrane~\cite{boulbitch.a:2002.a,dezerno.m:2003.a}.  A detailed
comparison of these processes requires, however, further analysis.

To summarize, we studied the interaction of a cylindrical colloidal
particle with a NI interface. We found that, when the particle is
close to the interface, the force on the particle is proportional to
the penetration depth $d/R$, and the amplitude of the force scales as
$R^2$. At larger penetrations, discontinuous configurational
transitions occur, related to the creation/annihilation of topological
defects.

\begin{acknowledgments}
  It is pleasure to thank M. Deserno, Yu. Reznikov, and A. Glushchenko
  for stimulating discussions.  DA acknowledges the support of the
  Alexander von Humboldt foundation.
\end{acknowledgments}

%\bibliography{journals,extra}

\end{document}